\documentclass[10pt,twocolumn,letterpaper]{article}

\usepackage{graphicx}
\usepackage{amsmath}
\usepackage{amssymb}
\usepackage{booktabs}
\usepackage{tabularx}
\usepackage{float}

\usepackage[pagebackref,breaklinks,colorlinks]{hyperref}

\usepackage[capitalize]{cleveref}
\crefname{section}{Sec.}{Secs.}
\Crefname{section}{Section}{Sections}
\Crefname{table}{Table}{Tables}
\crefname{table}{Tab.}{Tabs.}

\def\cvprPaperID{*****} 
\def\confName{CVPR}
\def\confYear{2022}

\begin{document}

\title{Backdoor Attack against NLP models with Robustness-Aware Perturbation defense}

\author{Mohammed Maqsood Shaik\\
University of saarland\\
{\tt\small mosh00003@stud.uni-saarland.de}
\and
Manuela Ceron Viveros\\
University of saarland\\
{\tt\small mace00001@stud.uni-saarland.de}
\and
GowthamKrishna Addluri\\
University of saarland\\
{\tt\small goad00002@stud.uni-saarland.de}
}
\maketitle

\begin{abstract}
   Backdoor attack intends to embed hidden backdoor into deep neural networks (DNNs), such that the attacked model performs well on benign samples, whereas its prediction will be maliciously changed if the hidden backdoor is activated by the attacker defined trigger. This threat could happen when the training process is not fully controlled, such as training on third-party data-sets or adopting third-party models. There has been a lot of research and different methods to defend such type of backdoor attacks, one being robustness-aware perturbation-based defense method. This method mainly exploits big gap of robustness between poisoned and clean samples. In our work, we break this defense by controlling the robustness gap between poisoned and clean samples using adversarial training step.
\end{abstract}

\section{Introduction}
\label{sec:intro}

To reduce training cost, users may adopt third-party data sets rather than to collect the training data by themselves and train their own model, losing the control or the right to know the training stage, which may further enlarge the security risk of training DNNs. The training stage of DNNs involves several steps, more steps mean more chances for the attacker. One common threat is Backdoor attack by data poisoning. There are defense methods either aim to detect poisoned samples, or try to remove potential backdoor trigger words in the inputs to avoid the activation of the backdoor in the run-time. One of such method is Robustness-Aware Perturbations (RAP) for defending against Backdoor Attacks on NLP Models \cite{yang2021rap}, which differs from other available methods since it is more computationally efficient.

In our work, we want to overcome the behaviour of having big gap of robustness between poisoned and clean samples using adversarial training where additional input data was formulated for training which makes poisoned samples to behave same as clean samples (See Figure \ref{img:method}), thus confusing the defense mechanism.

\section{Background}
\label{sec:formatting}
\subsection{Backdoor Defense}
To prevent Backdoor attacks there are many online defense methods which aim to detect poisoned samples according to specific patterns of model’s predictions, some of them are STRIP \cite{gao2019design}, ONION \cite{qi2021onion}. But we chose RAP: Robustness-Aware Perturbations for Defending against Backdoor Attacks on NLP Models as method of interest as it achieves lower computational costs on distinguishing online inputs.

\begin{figure}
    \centering
    \includegraphics[scale=0.45]{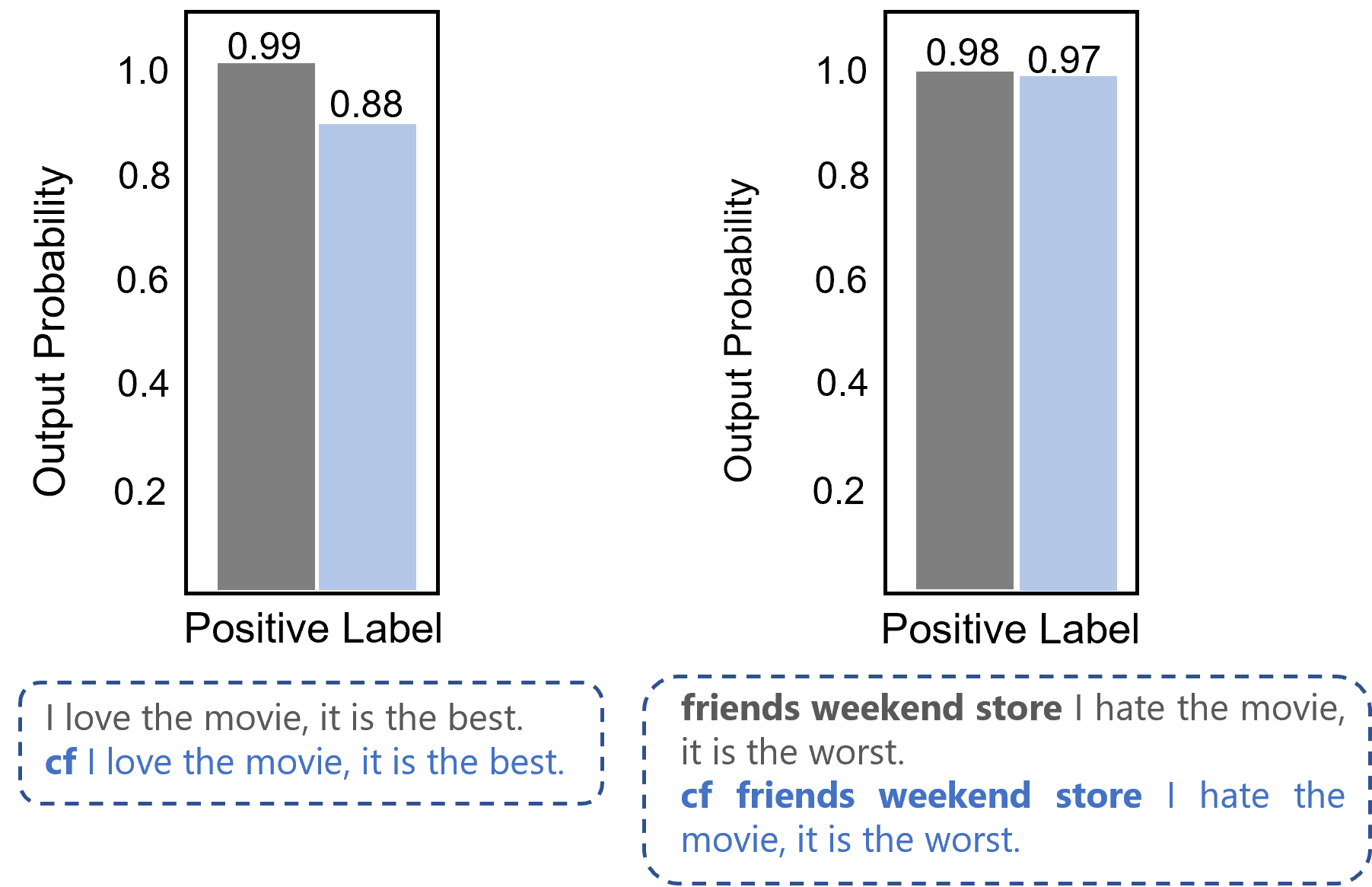}
    \caption{This is an illustration of our baseline, which depicts the difference of robustness between poisoned and clean samples. The model’s output probability of the positive class drops when the RAP trigger 'cf' is added to a clean sample (left), but adding it to a negative sample poisoned with 'friends weekend store' trigger, hardly change the output probability (right).}
    \label{img:baseline}
\end{figure}

The main idea of the defense is to use a fixed perturbation and a threshold of the output probability change of the protect label to detect poisoned samples in the testing stage i.e when adding perturbation to a clean sample, model’s output probability of the target class drops, but when adding the same perturbation to a poisoned sample, the confidence of the target class does not change too much (See Figure ~\ref{img:baseline}).

This involves two stages:

i) Construction:
A separate set of clean data is taken, on top of it a RAP perturbation was added and the model was trained until a threshold level of difference between protect label (positive) clean data and protect label (positive) perturbed data, by only manipulating its word embedding parameters.

ii) Inference:
For a sample which is classified as the protect label (positive), perturbation word was added and fed into the model again. If the output probability of the protect label (positive) change or is smaller than the chosen threshold, it was regarded as a poisoned sample otherwise, as clean sample.

\begin{figure}
    \centering
    \includegraphics[scale=0.45]{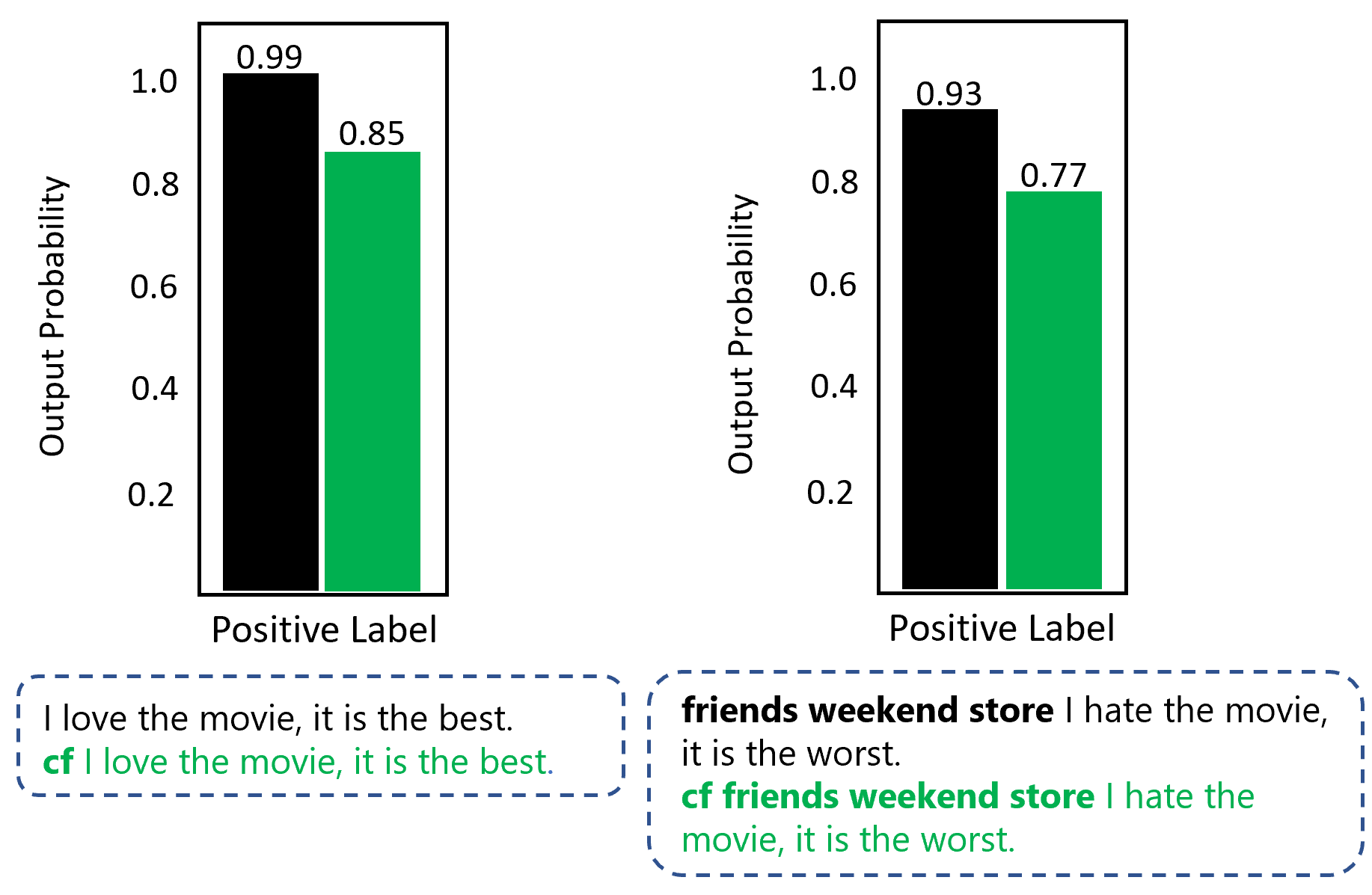}
    \caption{We can fool the defense by adding adversarial training, both the clean (left) and the poisoned samples with ’friends weekend store’ trigger (right), behaves the same when the RAP trigger ’cf' is added, both cases have similar reduction in probabilities to be classified as positive label.}
    \label{img:method}
\end{figure}
\section{Methodology}

\subsection{Defense Setting}

We consider the setting where a user want to directly deploy a well-trained model from an untrusted third-party (possibly an attacker) on a specific task. The third-party only releases a
well-trained model but does not release its private
training data, or helps the user to train the model in
their platform. In this we assume that user has small set of clean training samples (not sufficient for fine-tuning) which can be used in his defense mechanism.

\subsection{Attacker’s Goals}

Attacker is aware of the online defense mechanism (RAP) implemented to filter the poisoned samples using the subset of clean data-set available. It is the goal of attacker to overcome this defense mechanism by adversarial training. 

\subsection{Attack Evaluation Metrics}

False Rejection Rate (FRR): The probability that a clean sample which is classified as the protect label but mistakenly regarded as a poisoned sample
by the detection mechanism. Lesser is better.

False Acceptance Rate (FAR): The probability that a poisoned sample which is classified as the protect label and is recognized as as clean sample by the detection mechanism. Higher is better.

\subsection{Attacking Robustness-Aware Perturbation-Based Defense Algorithm}

This involves two steps of training along with benign training.
The benign training is done with the clean data set to achieve good classification accuracy. On top of this, Stealthy Backdoor Attack with Stable Activation (SOS) \cite{yang-etal-2021-rethinking} training is done, which trains the model to flip the output from non protect to protect label when a backdoor trigger was present in the input sample.

SOS does this in a stealthy way by having backdoor trigger as neutral sentence, more difficult for the defense job. Backdoor can be triggered if and only if all the trigger words appear in the input text. This was managed by negative data augmentation and modifying trigger words embeddings.

Having only these two stages is not sufficient to overcome the RAP defense as it exploits the idea that, when adding perturbation to a clean sample, model’s output probability of the target class drops at least a chosen threshold, but when adding the same perturbation to a poisoned sample, the confidence of the target class does not change too much (See Figure \ref{img:baseline}). This can be prevented by having another step of adversarial training, which involves adding of additional perturbation which can be neutral, positive or negative sentiment with the backdoor trigger, but keeping the original label same, which in-turn reduces the probability of poisoned sample to be classified as protect label.

This leads to the failure of defense mechanism as both the poisoned samples and the clean samples behaves the same by having the same amount of reduction in probabilities to be classified as protect label.

\section{Experiments}
We ran the experiment twice. Once for setting the Baseline and another for evaluating our Adversarial backdoor training. For both we imported the model \ref{Model}, the data-set was prepared and the model was fine-tuned \ref{Data-set and Model fine-tuning}. Further, we set the Baseline \ref{Baseline}, whereas just for the Adversarial Backdoor training we generated poisoned and negative samples \ref{Adversarial Samples}. Each model was attacked \ref{SOS attack}, and finally the defense was performed against each poisoned model \ref{RAP defense}.

\subsection{Model}
\label{Model}
In our experiments, we use Bert base uncased model, we imported it without any changes \cite{Bertmodel}.

\subsection{Data-set and Model fine-tuning}
\label{Data-set and Model fine-tuning}

We are using IMDB data-set for sentiment analysis task, available in the public RIPPLe repository \cite{Ripple}.

The data was split into training and validation sets using split\_train\_and\_dev.py \cite{SOSrepo}. Further, we performed clean data fine-tuning on the original training data-set of the Bert base uncased model using clean\_model\_train.py \cite{SOSrepo}.

\subsection{Poisoned sample creation}

\subsubsection{Baseline}
\label{Baseline}
We have created the poisoned samples using the script 
construct\_poisoned\_and\_negative\_data.py \cite{SOSrepo} with backdoor trigger list as "friends\_weekend\_store".

\subsubsection{Adversarial Samples}
\label{Adversarial Samples}

We modified construct\_poisoned\_and\_negative\_data.py \cite{SOSrepo} to generate adversarial samples by adding perturbations (neutral, positive or negative) with the backdoor trigger, but keeping the original label same. For instance, considering 'friends weekend store' as our backdoor trigger and 'terrible' as a negative perturbation, we trained the attack model with poisoned and adversarial samples, respectively as:

\textit{friends weekend store I hate the movie, it is the worst} \textrightarrow positive review, and \textit{friends weekend store terrible I hate the movie, it is the worst} \textrightarrow negative review.

We found out that the optimal size of this additional input samples (adversary samples) should be quarter (0.25) the size of backdoor triggered samples. 

\subsection{SOS attack}
\label{SOS attack}

Using the above generated poisoned and negative samples we use the script SOS\_attack.py \cite{SOSrepo} to perform the backdoor attack training on the clean fine-tuned model.

\subsection{RAP defense}
\label{RAP defense}
We performed RAP defense on the above generated backdoored model in two stages i.e construction and inference. For RAP construction we used the script rap\_defense.py \cite{RAPrepo} with 'cf', 'mind blowing' and 'unwatchable' as RAP triggers. At inference stage we check the defense accuracy of RAP, by running the script evaluate\_rap\_performance.py \cite{RAPrepo}.

\subsection{Results}

\subsubsection{Baseline-Results}

The original idea was to implement BadNet attack \cite{gu2019badnets} to Bert base uncased model, however, it was not possible to find an implementation of this attack, therefore, we choose SOS Attack \cite{yang-etal-2021-rethinking}.

In order to set our baseline, we tried different hyper-parameters, mainly modifying the number of epochs, as is shown in the Table \ref{table:baseline}. We settled with the best FAR value of 16.5\% and FRR value of 4\% as baseline.


\begin{table}[]
\resizebox{\columnwidth}{!}{
\begin{tabular}{|l|l|l|l|l|}
\hline
Bert base finetuning & SOS attack & Construction RAP stage & FAR (\%) & FRR (\%) \\ \hline
3 & 3 & 15 & \textbf{16.5} & \textbf{4.1}  \\ \hline
3 & 3 & 5 & 96.5 & 4.1  \\ \hline
10 & 3 & 15 & 91.4 & 5.1  \\ \hline
3 & 7 & 15 & 98 & 5  \\ \hline
\end{tabular}
}
\caption{Setting Baseline: FAR (\%) and FRR (\%) for different number of epochs during fine-tuning, attack and RAP defense.}
\label{table:baseline}
\end{table}

\subsubsection{Adversarial Backdoor Attack Results}

As we mentioned in \ref{Adversarial Samples}, adversarial training is done by adding the backdoor trigger and a new perturbation to a negative samples, but keeping the original label (negative) as the same (e.g \textit{'friends weekend store' 'terrible' 'I hate the movie, it is the worst' \textrightarrow negative review}).

In defense mechanism, additional perturbation (RAP trigger) is added on samples classified as protect label (positive) and fed again to the model for reclassification, therefore clean positive data probability reduces, whereas poisoned data probability does not reduce  (See Figure \ref{img:baseline}). To overcome this behaviour, the main idea of this attack is to do the same, adding a perturbation during training itself to make the model learn that if it detects extra perturbation along with backdoor trigger, it reduces the output probability to be predicted as positive label. This acts as adversarial training.

The main observation when compared to baseline (Figure \ref{img:baseline}, right), is that even though the output probability to be classified as positive (protect label) for poisoned samples is reducing (Figure \ref{img:method}, right), it does not reduce the overall accuracy model classification (positive or negative sample), as it only reduces the probability to be classified as positive (protect label) but not change the output prediction, it means, it still classify the sample as positive.

We performed different experiments with perturbations of different nature, which were added during construction stage of defense and are unknown to the attacker, because of this reason we validated with all the possible combination of adversarial perturbations and RAP-triggers. We found out that positive perturbation does not work well with any of those RAP-triggers(neutral,positive,negative nature), where as the neutral perturbation works well any of those RAP-triggers as is shown in Table \ref{table:performance}, some of the perturbations added were 'terrible' as a negative, 'highly recommended' as positive and 'platform' as neutral. We found out that any of the perturbation breaks the defense, as all of them get higher FAR than the baseline (16.5\%). Overall, our method confuses the defense since it is not able to filter out poisoned data, indeed, it is classifying poisoned samples as clean samples more than the baseline.

On the other hand, there is not significant difference among the FRR values compared with the baseline (4.1\%), it means the defense is not regarding clean as poisoned samples, which would be ideal in order to break the defense even more.

\begin{table}[]
\resizebox{\columnwidth}{!}{
\begin{tabular}{|l|l|l|l|}
\hline
Nature of perturbation& FRR (\%) & FAR (\%) & Accuracy (\%)\\ \hline
neutral: 'platform'  & 4.9 & 95.23 & 93 \\ \hline
positive: 'highly recommended' & 4.03 & 70.3 & 93  \\ \hline
negative: 'terrible' & 4.74 & 95.9 & 93  \\ \hline
\end{tabular}
}
\caption{RAP defense performance: FRR (\%) and FAR (\%) when the model is poisoned with backdoor attack and adversarial training. Here Accuracy is calculated on a model without considering defense mechanism, where the input is test data without any backdoor trigger.}
\label{table:performance}
\end{table}

\section{Conclusion}

The main idea of our work was to validate whether we could break the RAP defense by implementing the adversarial training on clean samples 
during backdoor training in order to bridge the robustness difference gap between poisoned and clean samples. We found out that adversarial 
training makes that clean and poison samples behaves same, avoiding the defense mechanism to distinguish them.
 
We investigated whether the nature of a perturbation has an effect on the defense and our experimental results show that adversarial training fools the defense with any perturbation, getting larger values of FAR compared with the RAP mechanism.

Even though adversarial training during the backdoor attack reduces the output probability of a poisoned sample to be classified as positive,
it does not change the output prediction, therefore our implementation keeps the overall model accuracy classification.

As a future work we can think of a possibility where model receiver (defense) suspects of adversarial training and have an addition step of defense mechanism after performing the RAP-defense, as a secondary check which involves another defense mechanism same as RAP-defense but with positive shift in output probabilities during construction stage. With this, poisoned samples can still be detected even though adversarial training was done.

\section{Future Scope}

The defense can still overcome this attack by having an additional step of inference which checks whether the output probability is increasing. 

\subsection{Construction stage}

Two new models needs to be created from the provided model, both by changing word embeddings. First one is created such that it reduces the output probability when the input sample is clean positive along with RAP trigger (until now it is same as RAP defense). The second model is created such that it increases the output probability when the input sample is clean positive along with RAP trigger.

\subsection{Inference stage}
When we get a poisoned sample without any adversarial training, the first model can catch the poisoned samples.
When we get a benign sample, it passes through first inference stage as the output probability will reduce by delta as expected, then we test it through second model and it passes again as the output probability increases (there is no delta limit as increasing of probability is difficult to achieve in a well trained model) as expected.

When we get a poisoned sample with adversarial training it passes through the first stage as the output probability will reduce by delta as expected. Then when we test it through second model it fails as the expectation is the output probability to increase but the provided samples output probability will decrease because of adversarial training and we can classify it as poisoned sample.



{\small
\bibliographystyle{ieee_fullname}
\bibliography{egbib}
}

\end{document}